# Fermi Level Engineering and Mechanical Properties of High Entropy Carbides

MD Hossain, S Lowum, T Borman, J-P Maria

Department of Materials Science and Engineering, The Pennsylvania State University, University Park, PA 16802, USA

**Abstract:**

Fermi level engineering and mechanical properties evolution in high entropy carbides are investigated by theoretical and experimental means. Massive elemental diversity in high entropy ceramics broadens the compositional space but imposes great challenges in composition selection and property investigation. We have utilized the valence electron concentration (VEC) descriptor to design and predict properties of high entropy carbides. The VEC regulates the Fermi energy and systematically alters the bonding characteristics of materials. As a result, mechanical properties evolve as function of the VEC. At VEC 8.4, the strong $\sigma$ bonding states stem from filled overlapping metal $d$ and carbon $p$ orbitals, which results in maximum resistance to shear deformation and highest hardness. Beyond or below the optimum VEC point of 8.4, mechanical response degrades due to filling or emptying of energy orbitals that facilitates shear deformation. Furthermore, the optimum VEC point can shift based on the constituent metals that formulate the high entropy carbide. Our analyses demonstrate strong correlation between calculated hardness and shear modulus. As an experimental complement, a set of high entropy carbides are synthesized, and mechanical properties investigated. The measured hardness follows theoretical predictions and the highest hardness of ~30 GPa is achieved at VEC 8.4. In contrast, hardness decreases by 50% when VEC is 9.4. Designing high entropy carbides based on VEC and understanding mechanical properties at an electronic level enables one to manipulate the composition spectrum to procure a desired mechanical response from a chemically disordered crystal.

**Introduction:**

Ultra-high temperature ceramics (UHTC) with melting temperatures greater than 3000 $^0$C are suitable for extreme environment applications: from nuclear reactors to hypersonic missile propulsion systems, leading edges and nosetips that encounter blazing temperatures during operation.[1,2] In addition to high melting temperature, UHTC materials need to satisfy several other vital criteria for extreme engineering applications, such as oxidation resistance, high temperature strength, high thermal conductivity, low thermal expansion, cheap and easy manufacturability, etc. There are few existing materials, mostly IVB and VB transition metal borides, carbides and nitrides, that possess this essential and intricate set of properties for extreme environment applications.[1,2] In recent years, with renewed interest in hypersonic technologies, there has been extensive drive to discover and expand the UHTC materials spectra.

High entropy carbides are a new class of UHTC materials that demonstrate exceptional thermal, electrical, mechanical and oxidation properties.[3–9] These chemically disordered crystals routinely exhibit high hardness compared to their constituent transition metal carbides.[6,8,9] Thermal conductivity measurements revealed extremely low and atypical thermal transport characteristics.[4,7] High temperature study demonstrated selective oxidation and improved creep resistance behavior.[3,5,10,11] It has been conjectured chemical disorder induces high temperature stability and solid solution strengthening responsible for enhanced mechanical properties.[9] However, these improved and anomalous properties along with their underlying mechanisms is yet to be understood on an atomic and electronic level. Furthermore, high entropy carbides span a huge composition spectrum with few compositions experimentally and theoretically explored to date.

One of the most widely accepted definitions of high entropy materials is an alloy composed of five or more principle elements in equimolar ratios.[12,13] Another definition extends the composition spectrum beyond equimolar ratios and includes alloys with five or more elements but concentration varies from 5-35 at.%.[12,13] Based on these definitions, there are thousands of high entropy carbide formulations possible when only considering IIIB, IVB, VB and VIB transition metals as constituents. Although these new compositions tremendously expand the materials palette and enrich the UHTC repository, it is a gargantuan task to explore these many compounds computationally or experimentally. If not all, a substantial number of materials needs to be investigated to make an informed decision on composition selection for any extreme engineering application. Another way to resolve this complex problem is by recording the physical, chemical and mechanical properties evolution with respect to a set of distinct or interdependent descriptors such that those descriptors can then be used to design a new material with predetermined physicochemical properties. This approach may be exceptionally beneficial for designing extreme environment materials where a combination of physical, chemical and mechanical properties are required.

Hence, we have used Fermi level engineering/valence electron concentration as a descriptor to design and predict mechanical properties of high entropy carbides. Fermi level engineering is a prevalent and powerful methodology that has been utilized to regulate magnetic, electronic, optical, thermal, and mechanical properties of diverse material systems, such as organic, inorganic semiconductors, oxides, carbides, nitrides, borides, etc.[14–22] At an electronic level, dopant introduction alters carrier concentration and mobility and, therefore, directly influences the electrical, optical and thermal properties to some degree. But a direct correspondence between electronic adjustment and mechanical properties evolution is much more complicated due to pronounced influence of micro- and macro-scale features, namely grain boundaries, precipitates, and dislocations, on material mechanical properties.

Previous studies showed that mechanical and elastic properties of transition metal carbides could be described based on number of valence electrons in the lattice.[17–19,23–27] Transition metal carbides, nitrides with valence electron concentration of 8.4 demonstrated highest hardness. The valence electron concentration controls electronic occupation and changes bonding characteristics. Accordingly, we have designed high entropy carbides based on valence electron concentration. The composition list is provided in Table 1 and is categorized into two series based on constituents that formulate high entropy carbides. Series 2 includes Sc,Y plus three transition metals from IVB,

VB and VIB, whereas Series 1 exclude both IIIB, Sc and Y in same composition. First principles calculations were performed to demonstrate Fermi level engineering and mechanical properties evolution. Subsequently, high entropy carbide thin films were synthesized and characterized, then mechanical properties transformations were compared with theoretical predictions.

**Experiment:**

*Calculation Method:* Density functional theory calculations were performed with Quantum ESPRESSO v6.2.[28]. To simulate a chemically disordered crystal, an 80-atom supercell which ensured an equiatomic five cation distribution in the cell was constructed. Special Quasirandom Structures (SQS), a tool in Alloy Theoretic Automated Toolkit (ATAT) package, was utlized to generate a supercell with randomly distributed cations.[29] A 3×3×3 k-point grid based on the Monkhorst-Pack scheme was used for structure relaxation and energy calculation, whereas electronic structure was generated by using 8×8×8 k-point grid. The plane wave cutoff was set to 120 Ry for all calculations. Elastic constants were calculated based on a strain-energy formulation where the following strain tensor expressions were used to distort a supercell:[17,30,31]

$$\epsilon_{tetr} = \frac{1}{3}\begin{pmatrix} -\delta & 0 & 0 \\ 0 & -\delta & 0 \\ 0 & 0 & 2\delta \end{pmatrix} \quad (1)$$

$$\epsilon_{orth} = \frac{1}{3}\begin{pmatrix} 0 & \delta & 0 \\ \delta & 0 & 0 \\ 0 & 0 & \delta^2 \end{pmatrix} \quad (2)$$

The final lattice vectors relate to the ground state supercell in accordance with $\boldsymbol{R'}=(1+\epsilon)\boldsymbol{R}$, where $\boldsymbol{R}$ and $\boldsymbol{R'}$ are the starting and final distorted lattice vectors, respectively. The three independent elastic constants ($C_{11}$, $C_{12}$, and $C_{44}$) were calculated by fitting strain vs. energy data with the following quadratic equations:

$$U_{tetr} = \frac{1}{3}(C_{11} - C_{12})\delta^2 \quad (3)$$

$$U_{orth} = 2C_{44}\delta^2 \quad (4)$$

Table 1: List of high entropy carbides and their respective valence electron concentration (VEC).

| Composition | Designation | VEC |
|---|---|---|
| Series 1 | | |
| (HfNbTaTiZr)C | HEC3 | 8.4 |
| (HfNbTaTiV)C | HEC4 | 8.6 |
| (NbTaTiVW)C | HEC5 | 9.0 |
| (HfWMoTaZr)C | HEC6 | 9.0 |
| (HfWTaTiZr)C | HEC7 | 8.6 |
| (HfWMoTiZr)C | HEC8 | 8.8 |
| (NbTaTiHfW)C | HEC9 | 8.8 |
| (HfMoWZrV)C | HEC10 | 9.0 |
| (ScHfTiZrTa)C | HEC1000 | 8.0 |
| (ScHfZrTaW)C | HEC1002 | 8.4 |
| VNbTaMoWC | HEC16 | 9.4 |
| Series 2 | | |
| (ScYHfWMo)C | HEC1003 | 8.4 |
| (ScYZrTiHf)C | HEC1004 | 7.6 |
| (ScYZrMoHf)C | HEC1005 | 8.0 |
| (ScYVNbTa)C | HEC1006 | 8.2 |
| (ScYTaWMo)C | HEC1007 | 8.6 |

The theoretical hardness ($H_v$) of covalently bonded crystals was calculated with the model developed by Chen *at el.* which expresses hardness as a function of shear modulus and bulk modulus.[32] Bulk modulus ($B$) was calculated by fitting the Birch

Murnaghan isothermal equation of state to the energy vs. volume data (15 data points). Shear modulus ($G$), Young's modulus ($E$) and Poisson's ratio ($\upsilon$) were calculated according to these given equations: [17,30,31]

$$H_v = 2(k^2 G)^{0.585} - 3 \qquad (5)$$

$$k = \frac{G}{B} \qquad (6)$$

$$G = \frac{3(C_{11} - C_{12}) + C_{44}}{5} \qquad (7)$$

$$E = \frac{9BG}{3B + G} \qquad (8)$$

$$\upsilon = \frac{1}{2}\left(1 - \frac{E}{3B}\right) \qquad (9)$$

High entropy carbides were selected based on valence election concentration and include IIIB, IVB, VB and VIB transition metals. All compositions are listed in Table 1.

*Thin Film Synthesis:* Thin films were synthesized with reactive high-power impulse magnetron sputtering (HiPIMS). HiPIMS is a relatively new magnetron sputtering technique which provides greater opportunity to control the plasma energy, as well as thin film microstructure and stoichiometry.[33] The HiPIMS was operated in a preprogrammed condition that maintained a fixed average power and pulse width of 250 W and 40 μs, respectively. The pulse frequency was self-regulated and conserved the average power. The discharge voltage was set to -700 V. A +20 V kick pulse applied after the main pulse potentially guided the ionic plasma to the substrate. Equiatomic metal alloy targets (99.5%, ACI Alloys) with reference to the listed compositions in Table 1 served as a metal source. Before each deposition, targets were presputtered for 2 mins with Ar ($7.5 \times 10^{-3}$ Torr) and 2 mins with a combination of Ar and $CH_4$ gas. The methane gas was the source of carbon. All samples were deposited on *c*-sapphire substrates at ~650 $^0$C. The deposition rate was calibrated to obtain approximately ~300 nm thin films.

*X-ray Diffraction:* X-ray diffraction measurements were performed with a Panalytical Empyrean X-ray diffractometer equipped with a monochromatic Cu *Kα* radiation source operated at 45 kV and 40 mA. The incident optic utilized a Bragg-Brentano geometry where a 4 mm mask, 1/8 degree divergent slit and ½ degree anti-scatter slit shaped the incident beam. The diffracted beam was collected with a PIXcel$^{3D}$ detector. The 0.4 radian Soller slit and ¼ degree anti-scatter slit directed the diffracted beam to the area detector.

*Scanning Electron Microscopy:* Microstructural changes as a function of composition were observed using field emission scanning electron microscopy (FE-SEM) performed on a Thermo Fisher Scientific Verios G4 UC. A 1 kV accelerating voltage and a ~2 mm working distance were used with the Through Lens Detector.

*Nanoindentation:* Thin film mechanical properties were measured with a Hysitron TI-900 nanoindenter with a load resolution of 1 nN. The maximum applied load was 500 μN, where loading and unloading cycles were 5 s with a 2 s hold at the maximum load. Each sample was indented in a 3×3 square grid pattern with indents 20 μm apart. The Berkovich indenter tip area function was calibrated with a standard fused silica specimen, whereas the tip-to-optic distance was calibrated with a standard polycarbonate sample. Sample hardness and reduced elastic modulus were obtained from the load vs. displacement data according to the protocol described by Oliver-Pharr.[34,35]

**Results and Discussions:**

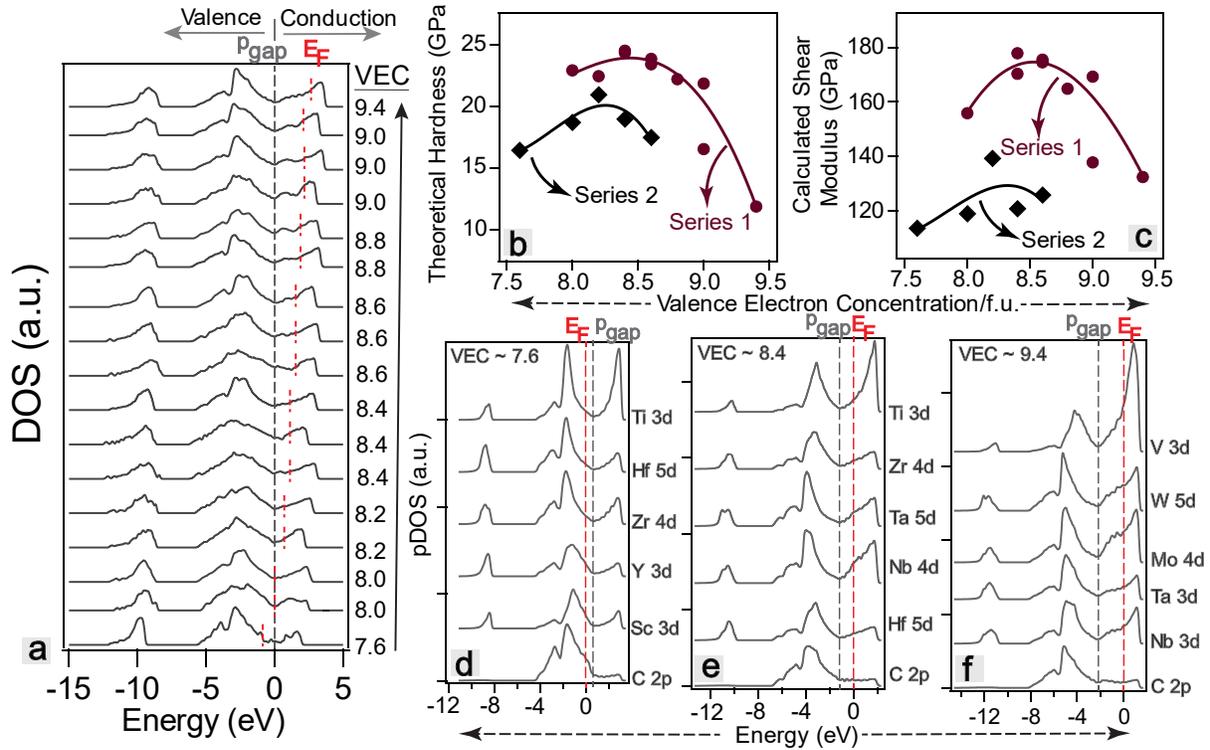

Figure 1: (a) Fermi level ($E_F$) engineering concept is demonstrated through electronic structure plot of high entropy carbides. Compositions were designed based on valence electron concentration (VEC) and electronic structures are organized with increasing VEC from bottom to top; the $E_F$ consistently shifts to the right with increasing VEC and modifies that material's bonding characteristics. The grey dotted line is the pseudogap ($p_{gap}$), which indicates the minima in the electronic structure and separates the valence and conduction bands. The $p_{gap}$ is the reference for a 0 energy state and the red dotted line represents the fermi level, $E_F$. (b) theoretical hardness of high entropy carbides is plotted as a function of the VEC (c) calculated shear moduli of high entropy carbides are plotted as a function of the VEC; compositions are listed as Series 1 and Series 2 in Table 1. Fermi level shifting as a function of the VEC changes occupation of metallic *d* orbitals, which is demonstrated through the partial density of states (pDOS) analysis of three high entropy carbide compositions with variable VEC: (d) pDOS of (ScYZrTiHf)C with VEC~7.6 showing the metallic Ti3*d*, Hf5*d*, Zr4*d*, Y3*d*, Sc3*d* and C2*p* orbitals (e) pDOS of (HfNbTaTiZr)C with VEC~8.4 showing the metallic Ti3*d*, Zr4*d*, Ta5*d*, Nb4*d*, Hf5*d* and C2*p* orbitals (f) pDOS of (VNbTaMoW)C with VEC~9.4 showing the metallic V3*d*, W5*d*, Mo4*d*, Ta5*d*, Nb3*d* and C2*p* orbitals.

Valence electron concentration controls the chemical reactivity of an element and shapes the bonding states of a compound. Assuming the crystal structure and symmetry remains intact, doping/alloying with different valence electron elements can presumably change the electronic occupation of a compound. By then tracking the Fermi energy, the ramifications of changes in valence electrons can be understood at an electronic level. Figure 1a shows the electronic structures of designed high entropy carbides that are organized from bottom to top with increasing valence electron concentration (VEC). The minima in the electronic structure represents the pseudogap ($p_{gap}$), which separates the pseudo valence and conduction bands. The Fermi level ($E_F$) and $p_{gap}$ align at VEC 8.0. Considering $p_{gap}$ is the reference state, the Fermi level position shifts as a function of the VEC as shown in Fig. 1a. Therefore, VEC regulates the occupied energy states and bonding in these covalent crystals. Consequently, material properties such as hardness that are correlated with bonding are anticipated to evolve with VEC. Theoretical hardness of these materials is plotted as a function of the VEC in Fig. 1b. Hardness data can be classified into two sets: for composition series 1 in Table 1 with VEC range 8.0-9.4, the maximum hardness results at a VEC of 8.4; whereas for composition series 2 with VEC range 7.6-8.6, maximum hardness emerges at a VEC of 8.2. The ability to predict hardness based on the VEC is highly convenient as one can use this as a guiding principle to design high entropy carbides with specific hardness values.

Macroscopic properties, such as hardness, also strongly depend on dislocation interactions with microstructural features and obstacles, such as other dislocations, precipitates, grain boundaries, second phases, etc. However, in the case of transition metal carbides, nitrides and carbonitrides, covalent bond strength determines mechanical properties.[18,36] In these covalent materials, bond breaking strength ultimately dictates dislocation nucleation and propagation. Hence, hardness could be appropriately coordinated with elastic properties such as shear modulus, which defines rigidity of a crystal against shape deformation. Calculated shear moduli for the high entropy carbides are plotted as a function of the VEC in Fig. 1c. Identical to the hardness vs. VEC trend, the shear moduli follow two distinct trends: for composition series 1 with VEC range 8.0-9.4, highest shear modulus results at VEC of 8.4; in contrast, for composition series 2 with VEC range 7.6-8.6, maximum shear modulus occurs at VEC of 8.2.

Previous studies reported that maximum hardness of the binary and ternary transition metal carbides, nitrides and carbonitrides results when the VEC of a compound is 8.4.[18,37,38] These studies concluded at this VEC the electronic energy bands linked to the σ bonding states are filled and maximize the resistance to shear deformation. The $\sigma$ bonding states originate from face-to-face overlapped metal $d$ and carbon $p$ orbitals and form strong bonds.[18,38,39] Hence, one can perceive the correlation between shear modulus maximization and enhanced hardness. The calculated hardness values are 10-14% of the shear moduli and broadly, during an indentation experiment materials deform plastically within 8-12% strain.[40] In contrast, binary and ternary compounds with VEC < 8.4 contain unoccupied σ bonding states and diminished mechanical responses. Whereas in compounds with VEC > 8.4, hardness and shear moduli decrease due to the increased occupation of metallic $d$-$t_{2g}$ orbitals.[38,39] These specific energy bands facilitate shear deformations via the $d$-$t_{2g}$ electronic orbital overlap along the <110> directions if these orbitals are partially or completely filled by electrons.[18,38,41] Our present high entropy composition series 1 (VEC range 8.0-9.4) follows this description of hardness enhancement via an electronic mechanism. In addition, we can further examine the systematic change in discrete electronic orbital occupation through partial density of states (pDOS) analysis of individual transition metal elements.

Data from pDOS analysis on three representative high entropy carbides is presented in Fig. 1d-f. First, the (ScYZrTiHf)C high entropy carbide (VEC~7.6), pDOS analysis shows the Fermi level is below $p_{gap}$ and there are empty metallic $d$ and C $2p$ orbitals in the valence band (Fig. 1d). Strong σ bonding states that stem from metal $e_g$ ($x^2$-$y^2$, $z^2$) and C $2p$ orbitals dominate the valence band region and, similar to the described binary and ternary systems, these bonding states are filled at VEC 8.4, as shown by pDOS analysis of (HfNbTaTiZr)C in Fig. 1e. Notably, at this VEC the Fermi level is slightly above $p_{gap}$, allowing an astute observer to argue that beyond $p_{gap}$ one starts filling weak σ bonding states ($d$-$t_{2g}$ orbitals) between next nearest neighbor metal atoms. But through first principle calculations, Jhi *et al.* demonstrated strong $σ$ bonding states extend beyond $p_{gap}$ and are filled at VEC 8.4.[18,39] The bonding characteristics change to a great extent with further increase in VEC. As an example, the (VNbTaMoW)C high entropy carbide (VEC~9.4) pDOS analysis reveals a large amount of metallic $t_{2g}$ ($xy$, $yz$, $zx$) orbital occupation with little to no change of C$2p$ orbital occupation (Fig. 1f). We can agree high entropy carbides have similar electronic characteristics as binary and ternary carbides, nitrides and carbonitrides. However, the greater flexibility in composition enables one to explore uncharted territories of material properties engineering which are simply inaccessible in conventional alloy design processes. For instance, an additional hardness and shear moduli trend are observed in Fig. 1b and c, where the highest hardness and shear modulus are derived from (ScYVNbTa)C with a VEC~ 8.2 (see Table 1 composition series 2).

For this composition series, the optimum VEC point (i.e. the VEC corresponding to the highest hardness) not only shifts to a value < 8.4 , but the calculated hardness is also substantially lower. Let us consider two high entropy carbides−(HfNbTaTiZr)C in series 1 and (ScYHfWMo)C in series 2−that have the same VEC of 8.4; the calculated hardness values are 24 GPa and 18 GPa, respectively. This illustrates that a VEC~8.4 is not a unique value that can arbitrarily be used to optimize high entropy carbide mechanical properties. Rather, constituent transition metal elements potentially dictate the optimum VEC point. Closely inspecting compositions listed in Table 1 reveals all high entropy carbides in series 2 accommodate Sc and Y (IIIB transition metals) in the disordered lattice. Both cations have three valence electrons and locally form a relatively weak covalent bond with carbon (valence electron~4) due to the shortage of one valence electron. Consequently, compositions like (ScYHfWMo)C with VEC~8.4 produce empty σ bonding states in 40% of lattice sites. In contrast, (HfNbTaTiZr)C, which also has VEC~8.4 but only IVB and VB transition metals in the lattice, possesses sufficient valence electrons to promote globally strong σ bonding states. We then conclude chemically disordered five component high entropy carbides designed based on two IIIB (Sc,Y) plus three IVB, VB and VIB transition metals would produce a different hardness trend compared to five component high entropy compositions designed based on IVB, VB and VIB transition metals. The latter constituents mix produces a hardness trend identical to previously studied binary and ternary systems, where maximum hardness is acquired at VEC 8.4.

At this point we will move on to describe other elastic properties of high entropy carbides based on the VEC descriptor. Fig. 2a shows the bulk moduli of high entropy carbides that are plotted as a function of the VEC. Bulk modulus linearly increases with increasing VEC and three linear trend lines fit the listed compositions in Table 1 (Fig. 2a). The bulk modulus increases by 65% when VEC is boosted by ~2.0/f.u. The calculated bulk moduli of binary transition metal (IIIB, IVB, VB, VIB) carbides shows an analogous linear trend (Fig. 2b). Here we assumed the rock salt phase for all binary compositions to compare with high entropy systems.

First, we will explain the VEC vs. bulk modulus trend at an atomic level. The periodic properties of elements determine that covalent radius decreases with increasing VEC in the same period. Approximately, a material's bulk modulus ($K$) can be expressed as $K \propto r_0^{-n-3}$, where $r_0$ is an element's atomic radius and the value of n depends on the material's bonding characteristics. Transition metal carbides are categorized to be a unique set of materials with a combination of covalent, ionic and metallic bonding; the high entropy carbides demonstrate similar bonding characteristics. Hence, if one considers n as constant, the bulk moduli trend in Fig. 1a could be explained by the decrease in average atomic radius of the chemically disordered crystal with increasing VEC, which therefore causes an increase in the bulk modulus.

Furthermore, the observed VEC vs. bulk modulus trend can also be explained at an electronic level. In a previous study, we demonstrated that lattice parameter and crystal volume linearly decrease with increasing VEC of high entropy carbides.[42] A steady inflation in VEC means a higher number of electrons are being constrained in a smaller volume and, consequently, lattice compressibility decreases and bulk modulus increases. We can test this hypothesis by evaluating the number of electronic states ($N_f$) at the Fermi level. The $N_f$ values for high entropy and binary carbides are plotted as a function of VEC in Fig. 2 c and d, respectively. With increasing VEC, $N_f$ increases linearly; particularly, three linear trend lines fit the high entropy compositions listed in Table 1, which mimics the VEC vs. bulk moduli trend (three linear lines fit the data) in Fig. 2a. Previous studies of rock salt binary and ternary transition metal carbides, nitrides and carbonitrides demonstrated similar VEC vs. bulk modulus characteristics.[38,43] In addition, designing compositions based on VEC to engineer mechanical properties has been reported for other materials and crystal systems, such as layered hexagonal ($P6_3/mmc$) $M_2AC$ ($M$=Ti, Zr, Hf, V, Nb, Ta, Cr, Mo, W etc. A= Al, Si, P, Ge, Ga, Sn etc.) MAX phases and cubic ($Pm\bar{3}m$) perovskite $R$Rh$_3$B ($R$=Y, Zr, Nb) phases, orthorhombic transition metal monoborides, all of which demonstrated an analogous correlation between VEC and bulk modulus.[44–47] This suggests VEC-based formulations could be extended to other high entropy ceramics that potentially stabilize in a cubic or less symmetric crystal system with heterogenous bonding, such as high entropy oxides, nitrides, borides, perovskites, or chalcogenides. Although this may be true, we also must acknowledge that the $N_f$ vs. VEC plot for both binary and high entropy systems (Fig. 2c, d) shows outlier data points (i.e. points that do not follow the linear trend) when the tabulated compositions have a VEC < 8.0. We speculate this deviation led to a wider distribution in the bulk modulus for high entropy systems at a fixed VEC.

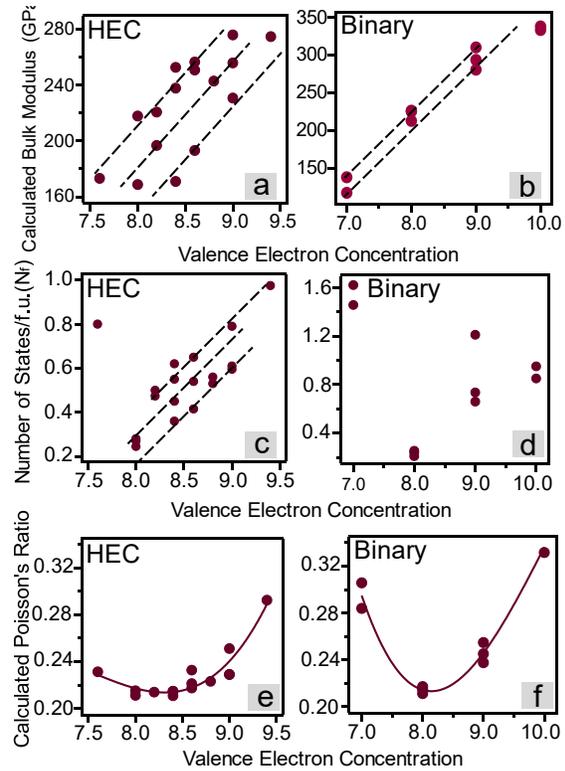

Figure 2: (a) Calculated bulk modulus of high entropy carbides (b) calculated bulk modulus of binary carbides that serve as constituents for the high entropy systems (c) the number of electronic states at the Fermi level ($N_f$) of high entropy carbides (d) the number of electronic states at the Fermi level ($N_f$) of binary carbides (d) Poisson's ratio (v) of high entropy carbides (e) Poisson's ratio (v) of binary transition metal carbides. All data are presented with respect to a composition's valence electron concentration; high entropy carbides and respective VEC values are listed in Table 1. Designated binary carbides and VEC values are: ScC, YC–VEC~7.0; TiC, ZrC, HfC–VEC~8.0; VC, TaC, NbC–VEC~9.0; MoC, WC–VEC~10.0. All binaries were assumed to be in rock salt phase. For high entropy carbides, three separate trend lines are drawn to demonstrate the linear fit to the bulk modulus and $N_f$ data, whereas a 4$^{th}$ degree polynomial was used to fit the Poisson's ratio.

A few other elastic properties of high entropy carbides can be illustrated with the VEC descriptor. As an example, Poisson's ratio (ν) of high entropy carbides and binary carbides is presented in Fig. 2e and f, respectively. Each materials system reveals a strong correlation between bonding states and Poisson's ratio evolution. As described earlier, the σ bonding states are filled at VEC~8.4, resulting in the lowest Poisson's ratio, whereas switching the Fermi level above or below that VEC point increases the Poisson's ratio. As the shear modulus is embedded in the Poisson's ratio formula (see calculation methods), one can determine the observed Poisson's ratio trend is a consequence of Fermi level engineering. The binary carbides show the lowest Poisson's ratio at VEC 8.0, as the optimum VEC point (~8.4) is inaccessible.

To validate the theoretical predictions, mechanical properties of a set of high entropy carbides were investigated experimentally. The compositions were selected from series 1, as outlined in Table 1, and enclose the VEC spectrum 8.0-9.4. X-ray diffraction patterns of the synthesized thin films are presented in Fig. 3a. All high entropy compositions result a single-phase rock salt structure, despite many constituents (Sc, Mo, W, V) in the mix prefer to form a non-cubic carbide structure. The rock salt high entropy phase may not be globally stable, but high effective temperature of plasma (~$10^5$ K) and low substrate temperatures (~650 $^0$C) introduce a drastic quenching effect (~$10^{12}$ K/s) and potentially stabilize a metastable high temperature rock salt phase at ambient conditions.[48] In a previous study, we have demonstrated that at high temperature configurational entropy dominates the free energy landscape and stabilizes single-phase high entropy carbides rather than any thermodynamically stable second phase or precipitate.[42]

We further examined the microstructural evolution of selected compositions; the scanning electron microscopy (SEM) images are presented in Fig. 3 b-i. Surface topographic images reveal carbide grains are nanocrystalline in nature, with grain size varying between ~10-100 nm. No apparent qualitative correlation between VEC and sample topography could be seen; all carbide specimens accommodate regular equiaxed, triangular, spherical or irregular grain morphologies. Noticeably, HiPIMS produces significantly smaller carbide grains compared to the RF and DC sputtered materials that were synthesized at comparable growth conditions.[7,49,50]

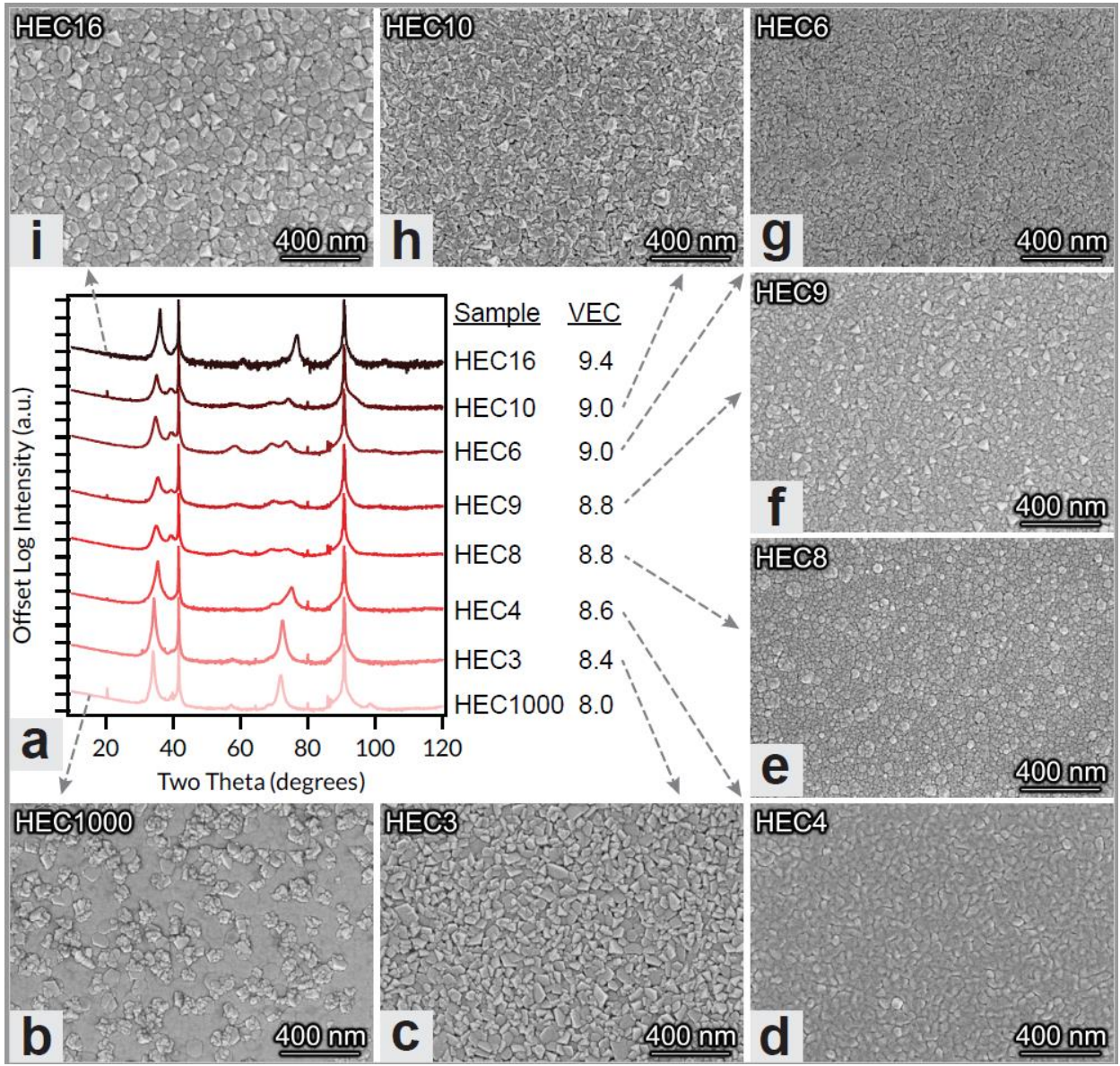

Figure 3: (a) X-ray diffraction pattern of high entropy carbides organized with increasing VEC from bottom to top, (b-i) scanning electron microscopy images of high entropy carbides show the microstructural evolution for different compositions; the selected high entropy compositions and respective VEC values are listed in Table. 1

Finally, thin film mechanical properties were investigated by nanoindentation. The measured hardness values are plotted as a function of the VEC in Fig. 4. In composition series 1 (VEC 8.0-9.4), the (HfNbTaTiZr)C with VEC 8.4 shows the maximum hardness of ~30 GPa and hardness reduces by 50% at the terminal VEC point of 9.4. Evidently, the measured hardness trend closely follows the theoretical prediction. Whereas at an electronic level theoretical hardness prediction is straightforward with its dependency on elastic modulus described earlier, empirical mechanical properties evolution is much more complex due to the crucial influence of micro- and macro-scale features on hardness.

One such factor is carbon stoichiometry; carbon vacancies significantly affect transition metal carbide mechanical properties. Frequently, hardness decreases linearly with vacancy concentration [49], but group VB transition metal carbides exhibit anomalous hardening behavior where certain percentages of carbon vacancies rather than softening the crystal improve mechanical properties.[37,51] Thus, to compare hardness of different high entropy carbides on the same scale, a perfect stoichiometric crystal where adverse or beneficial effects from a sub-stoichiometric lattice would not obfuscate property description is necessary. To ensure such crystals, thin films were synthesized by reactive high-power impulse magnetron sputtering (R-HiPIMS). This technique facilitates stoichiometric crystal growth that is elusive in reactive DC or RF sputtering of transition metal carbides.[52,53] The nanoindentation hardness data presented in Fig. 4 is representative of select stoichiometric crystals. Hence, we can discard any undue vacancy-related contribution to the observed hardness trend.

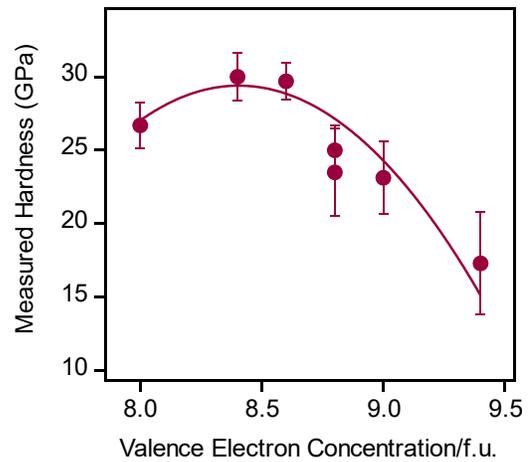

Figure 4: Nanoindentation hardness of high entropy carbides is plotted as function of the VEC.

Furthermore, microstructure-based hardening or domain hardening plays a pivotal role in transition metal carbides. The short and long range ordering of vacancies and presence of minor secondary phases−$Nb_6C_5$, $Ta_6C_5$ etc.−in rock salt transition metal carbide matrices frequently strengthens the parent crystal.[54–56] Any such secondary phases are nonexistent, as evident from the XRD patterns in Fig. 3a. Hence, we can conclude the experimental hardness trend originates due to changes in the materials' electronic occupation as a function of the VEC.

**Conclusion:**

Valence electron concentration (VEC) is a comprehensive descriptor that predicts mechanical properties of high entropy carbides. Calculated hardness, shear modulus and Poisson's ratio strongly correlates with a material's bonding: completely filled, strong $\sigma$ bonding states at VEC 8.4 results in the highest hardness and elastic properties. The optimum VEC point ~8.4 depends on the constituents of high entropy carbides and, as such, it can shift to a lower or higher VEC value. The bulk modulus increases linearly with VEC and correlates with the number of electronic states at the Fermi level. With increasing electronic population at the Fermi level, compressibility decreases. Selected high entropy carbide samples were synthesized with high power impulse magnetron sputtering (HiPIMS). All samples produced a single-phase rock salt structure with carbide grains approximately 10-100 nm. The measured hardness follows theoretical predictions; accordingly, high entropy carbides with VEC 8.4 show the highest hardness. Fermi level engineering and the VEC concept demonstrate a great potential for filtering high entropy carbides from a vast compositional space in order to obtain desired mechanical, thermal, and electrical properties. High entropy lattices accommodate diverse constituents without any phase separation. Consequently, one can design a set of high entropy carbides with an identical VEC, meaning the Fermi level is fixed. This empowers one to explore the constituent effects on the property evolution, as VEC is constant.

**Acknowledgement:** This research is funded by the U.S. Office of Naval Research Multidisciplinary University Research Initiative (MURI) program under Grant No. N00014-15-1-2863. This work further used the Extreme Science and Engineering Discovery Environment (XSEDE), Texas Advanced Computing Center (TACC) Stampede2 supercomputer resources which is supported by National Science Foundation grant numbers TG-DMR180016, TG-DMR180021 and TG-DMR170083. Computations for this research were partially performed at the Pennsylvania State University's Institute for Computational and Data Sciences' Roar supercomputer. Authors acknowledge Materials Characterization Laboratory (MCL) at Pennsylvania State University for SEM and nanoindentation characterizations. Any opinions, findings, and conclusions or recommendations expressed in this material are those of the authors and do not necessarily reflect the views of the Office of Naval Research or National Science Foundation.

**References:**

[1]  E. Wuchina, E. Opila, M. Opeka, W. Fahrenholtz, I. Talmy, UHTCs: Ultra-high temperature ceramic materials for extreme environment aplications, Electrochem. Soc. Interface. (2007) 30–36.

[2]  W.G. Fahrenholtz, E.J. Wuchina, W.E. Lee, Y. Zhou, Ultra-High Temperature Ceramics: Materials for Extreme Environment Applications, John Wiley & Sons, 2014. https://doi.org/10.1002/9781118700853.

[3]  X. Han, V. Girman, R. Sedlak, J. Dusza, E.G. Castle, Y. Wang, M. Reece, C. Zhang, Improved creep resistance of high entropy transition metal carbides, J. Eur. Ceram. Soc. 40 (2020) 2709–2715. https://doi.org/10.1016/j.jeurceramsoc.2019.12.036.

[4]  X. Yan, L. Constantin, Y. Lu, J.F. Silvain, M. Nastasi, B. Cui, $(Hf_{0.2}Zr_{0.2}Ta_{0.2}Nb_{0.2}Ti_{0.2})C$ high-entropy ceramics with low thermal conductivity, J. Am. Ceram. Soc. 101 (2018) 4486–4491. https://doi.org/10.1111/jace.15779.

[5]  Y. Tan, C. Chen, S. Li, X. Han, J. Xue, T. Liu, X. Zhou, H. Zhang, Oxidation behaviours of high-entropy transition metal carbides in 1200 °C water vapor, J. Alloys Compd. 816 (2020) 152523. https://doi.org/10.1016/j.jallcom.2019.152523.

[6]  T.J. Harrington, J. Gild, P. Sarker, C. Toher, C.M. Rost, O.F. Dippo, C. McElfresh, K. Kaufmann, E. Marin, L. Borowski, P.E. Hopkins, J. Luo, S. Curtarolo, D.W. Brenner, K.S. Vecchio, Phase stability and mechanical properties of novel high entropy transition metal carbides, Acta Mater. 166 (2019) 271–280. https://doi.org/10.1016/j.actamat.2018.12.054.

[7]  C.M. Rost, T. Borman, M.D. Hossain, M. Lim, K.F. Quiambao-Tomko, J.A. Tomko, D.W. Brenner, J.P. Maria, P.E. Hopkins, Electron and phonon thermal conductivity in high entropy carbides with variable carbon content, Acta Mater. 196 (2020) 231–239. https://doi.org/10.1016/j.actamat.2020.06.005.

[8]  P. Sarker, T. Harrington, C. Toher, C. Oses, M. Samiee, J.P. Maria, D.W. Brenner, K.S. Vecchio, S. Curtarolo, High-entropy high-hardness metal carbides discovered by entropy descriptors, Nat. Commun. 9 (2018). https://doi.org/10.1038/s41467-018-07160-7.

[9]  E. Castle, T. Csanádi, S. Grasso, J. Dusza, M. Reece, Processing and properties of high- entropy ultra-high temperature carbides, Sci. Rep. 8 (2018). https://doi.org/10.1038/s41598-018-26827-1.

[10]  L. Backman, J. Gild, J. Luo, E.J. Opila, Part I: Theoretical predictions of preferential oxidation in refractory high entropy materials, Acta Mater. 197 (2020) 20–27. https://doi.org/10.1016/j.actamat.2020.07.003.

[11]  L. Backman, J. Gild, J. Luo, E.J. Opila, Part II: Experimental verification of computationally predicted preferential oxidation of refractory high entropy ultra-high temperature ceramics, Acta Mater. 197 (2020) 81–90. https://doi.org/10.1016/j.actamat.2020.07.004.

[12]  D.B. Miracle, O.N. Senkov, A critical review of high entropy alloys and related concepts, Acta Mater. 122 (2017) 448–511. https://doi.org/10.1016/j.actamat.2016.08.081.

[13]  J.W. Yeh, S.K. Chen, S.J. Lin, J.Y. Gan, T.S. Chin, T.T. Shun, C.H. Tsau, S.Y. Chang, Nanostructured high-entropy alloys with multiple principal elements: Novel alloy design concepts and outcomes, Adv. Eng. Mater. 6 (2004) 299-303+274. https://doi.org/10.1002/adem.200300567.

[14]  Z. Zhang, S.A. Lindley, D. Guevarra, K. Kan, A. Shinde, J.M. Gregoire, W. Han, E. Xie, J.A. Haber, J.K. Cooper, Fermi Level Engineering of Passivation and Electron Transport Materials for p-Type $CuBi_2O_4$ Employing a High-Throughput Methodology, Adv. Funct. Mater. 30 (2020) 1–12. https://doi.org/10.1002/adfm.202000948.

[15]  A. Sciuto, A. La Magna, G.G.N. Angilella, R. Pucci, G. Greco, F. Roccaforte, F. Giannazzo, I. Deretzis, Extensive Fermi-Level Engineering for Graphene through the Interaction with Aluminum Nitrides and Oxides, Phys. Status Solidi - Rapid Res. Lett. 14 (2020) 1–6. https://doi.org/10.1002/pssr.201900399.

[16]  V.A. Gubanov, A.L. Ivanovsky, V.P. Zhukov, Electronic structure of refractory carbides and nitrides, Cambridge University Press, 1994. https://www.cambridge.org/us/academic/subjects/engineering/materials-science/electronic-structure-refractory-carbides-and-nitrides?format=HB&isbn=9780521418850 (accessed October 27, 2018).

[17]  D.G. Sangiovanni, V. Chirita, L. Hultman, Electronic mechanism for toughness enhancement in $Ti_xM_{1-x}N$ M=Mo and W), Phys. Rev. B - Condens. Matter Mater. Phys. 81 (2010) 1–7. https://doi.org/10.1103/PhysRevB.81.104107.

[18]  S.H. Jhi, J. Ihm, S.G. Loule, M.L. Cohen, Electronic mechanism of hardness enhancement in transition-metal carbonitrides, Nature. 399 (1999) 132–134. https://doi.org/10.1038/20148.

[19]  Y. Liang, Z. Gao, P. Qin, L. Gao, C. Tang, The mechanism of anomalous hardening in transition-metal monoborides, Nanoscale. 9 (2017) 9112–9118. https://doi.org/10.1039/c7nr02377d.

[20]  R. Friedrich, B. Kersting, J. Kortus, Fermi level engineering in organic semiconductors for controlled manufacturing of charge and spin transfer materials, Phys. Rev. B - Condens. Matter Mater. Phys. 88 (2013) 1–6. https://doi.org/10.1103/PhysRevB.88.155327.

[21]  M.J. Reed, F.E. Arkun, E.A. Berkman, N.A. Elmasry, J. Zavada, M.O. Luen, M.L. Reed, S.M. Bedair, Effect of doping on the magnetic properties of GaMnN: Fermi level engineering, Appl. Phys. Lett. 86 (2005) 1–3. https://doi.org/10.1063/1.1881786.

[22]  E.E. Perry, Fermi Level Engineering of the Perovskite and Electron Transport Layer Interface Through Charge Transfer Doping, 2018. https://search.proquest.com/docview/2135377682?accountid=13042%0Ahttp://oxfordsfx.hosted.exlibrisgro

up.com/oxford?url_ver=Z39.88-2004&rft_val_fmt=info:ofi/fmt:kev:mtx:dissertation&genre=dissertations+%26+theses&sid=ProQ:ProQuest+Dissertations+%26+Theses+G.

[23] R.F. Zhang, S.H. Sheng, S. Veprek, Origin of different plastic resistance of transition metal nitrides and carbides: Stiffer yet softer, Scr. Mater. 68 (2013) 913–916. https://doi.org/10.1016/j.scriptamat.2013.01.040.

[24] K. Han, G. Lin, C. Dong, K. Tai, X. Jiang, Experimental study on atomic-scale strengthening mechanism of the IVB transition-metal nitrides, J. Alloys Compd. 696 (2017) 572–579. https://doi.org/10.1016/j.jallcom.2016.11.329.

[25] S.H. Jhi, S.G. Louie, M.L. Cohen, J.W. Morris, Mechanical instability and ideal shear strength of transition metal carbides and nitrides, Phys. Rev. Lett. 87 (2001) 75503-1-75503–4. https://doi.org/10.1103/PhysRevLett.87.075503.

[26] D.G. Sangiovanni, L. Hultman, V. Chirita, Supertoughening in B1 transition metal nitride alloys by increased valence electron concentration, Acta Mater. 59 (2011) 2121–2134. https://doi.org/10.1016/j.actamat.2010.12.013.

[27] J.J. Gilman, Physical chemistry of intrinsic hardness, 1996. https://doi.org/10.1016/0921-5093(95)10116-0.

[28] P. Giannozzi, S. Baroni, N. Bonini, M. Calandra, R. Car, C. Cavazzoni, D. Ceresoli, G.L. Chiarotti, M. Cococcioni, I. Dabo, A. Dal Corso, S. de Gironcoli, S. Fabris, G. Fratesi, R. Gebauer, U. Gerstmann, C. Gougoussis, A. Kokalj, M. Lazzeri, L. Martin-Samos, N. Marzari, F. Mauri, R. Mazzarello, S. Paolini, A. Pasquarello, L. Paulatto, C. Sbraccia, S. Scandolo, G. Sclauzero, A.P. Seitsonen, A. Smogunov, P. Umari, R.M. Wentzcovitch, QUANTUM ESPRESSO: a modular and open-source software project for quantum simulations of materials, J. Phys. Condens. Matter. 21 (2009) 395502. https://doi.org/10.1088/0953-8984/21/39/395502.

[29] A. Van De Walle, M. Asta, G. Cederb, The Alloy Theoretic Automated Toolkit: A User Guide, 2002.

[30] Z. Wu, X.J..-S. pd. Chen, V. V. Struzhkin, R.E. Cohen, Trends in elasticity and electronic structure of transition-metal nitrides and carbides from first principles, Phys. Rev. B - Condens. Matter Mater. Phys. 71 (2005) 1–5. https://doi.org/10.1103/PhysRevB.71.214103.

[31] J. Yang, F. Gao, Hardness calculations of 5d transition metal monocarbides with tungsten carbide structure, Phys. Status Solidi Basic Res. 247 (2010) 2161–2167. https://doi.org/10.1002/pssb.201046127.

[32] X.Q. Chen, H. Niu, D. Li, Y. Li, Modeling hardness of polycrystalline materials and bulk metallic glasses, Intermetallics. 19 (2011) 1275–1281. https://doi.org/10.1016/j.intermet.2011.03.026.

[33] J. Keraudy, R.P.B. Viloan, M.A. Raadu, N. Brenning, D. Lundin, U. Helmersson, Bipolar HiPIMS for tailoring ion energies in thin film deposition, Surf. Coatings Technol. 359 (2019) 433–437. https://doi.org/10.1016/j.surfcoat.2018.12.090.

[34] G.M. Pharr, W.C. Oliver, Measurement of thin film mechanical properties using nanoindentation, MRS Bull. 17 (1992) 28–33. https://doi.org/10.1557/S0883769400041634.

[35] W.C. Oliver, G.M. Pharr, An improved technique for determining hardness and elastic modulus using load and displacement sensing indentation experiments, J. Mater. Res. 7 (1992).

[36] A. Kelly, N.H. Macmillan, Strong Solids, 3rd ed., Oxford [Oxfordshire] : Clarendon Press, 1986.

[37] H. Holleck, Material selection for hard coatings, J. Vac. Sci. Technol. A-Vacuum Surfaces Film. 4 (1986) 2661–2669. https://doi.org/10.1116/1.573700.

[38] K. Balasubramanian, S. V. Khare, D. Gall, Valence electron concentration as an indicator for mechanical properties in rocksalt structure nitrides, carbides and carbonitrides, Acta Mater. 152 (2018) 175–185. https://doi.org/10.1016/j.actamat.2018.04.033.

[39] S.-H. Jhi, J. Ihm, Electronic structure and structural stability of $TiC_xN_{1-x}$ alloys, 1997.

[40] V. Richter, A. Beger, E. Wolfb, Characterisation and wear behaviour of TiN-and $TiC_xN_{1-x}$-coated cermets, Mater. Sci. 209 (2000) 353–357.

[41] K. Balasubramanian, S. Khare, D. Gall, Vacancy-induced mechanical stabilization of cubic tungsten nitride, Phys. Rev. B. 94 (2016) 36–38. https://doi.org/10.1103/PhysRevB.94.174111.

[42] M.D. Hossain, T. Borman, C. Oses, C. Toher, S. Curtarolo, D. Brenner, J.-P. Maria, Entropy Landscapping of High Entropy Carbides, (2020).

[43] E.I. Isaev, S.I. Simak, I.A. Abrikosov, R. Ahuja, Y.K. Vekilov, M.I. Katsnelson, A.I. Lichtenstein, B. Johansson, Phonon related properties of transition metals, their carbides, and nitrides: A first-principles study, J. Appl. Phys. 101 (2007). https://doi.org/10.1063/1.2747230.

[44] J.M. Schneider, D. Music, Z. Sun, Effect of the valence electron concentration on the bulk modulus and chemical bonding in Ta 2AC and Zr 2AC (A=Al, Si, and P), J. Appl. Phys. 97 (2005) 1–4. https://doi.org/10.1063/1.1861504.

[45] Z. Sun, D. Music, R. Ahuja, S. Li, J.M. Schneider, Bonding and classification of nanolayered ternary carbides, Phys. Rev. B - Condens. Matter Mater. Phys. 70 (2004) 1–3. https://doi.org/10.1103/PhysRevB.70.092102.

[46] Z. Sun, R. Ahuja, S. Li, J.M. Schneider, Structure and bulk modulus of $M_2AlC$ (M=Ti, V, and Cr), Appl. Phys. Lett. 83 (2003) 899–901. https://doi.org/10.1063/1.1599038.

[47] D. Music, J.M. Schneider, Influence of valence electron concentration on elastic properties of $RRh_3B$ (R=Y, Zr, and Nb), Appl. Phys. Lett. 89 (2006) 1–4. https://doi.org/10.1063/1.2356991.

[48] G.N. Kotsonis, C.M. Rost, D.T. Harris, J.P. Maria, Epitaxial entropy-stabilized oxides: Growth of chemically diverse phases via kinetic bombardment, MRS Commun. 8 (2018) 1371–1377. https://doi.org/10.1557/mrc.2018.184.

[49] M.D. Hossain, T. Borman, A. Kumar, X. Chen, A. Khosravani, S. Kalidindi, E. Paisley, M. Esters, C. Oses, C. Toher, S. Curtarolo, J.M. LeBeau, D. Brenner, J.-P. Maria, Carbon Stoichiometry and Mechanical Properties of High Entropy Carbide, Acta Mater. (2020).

[50] P. Malinovskis, S. Fritze, L. Riekehr, L. von Fieandt, J. Cedervall, D. Rehnlund, L. Nyholm, E. Lewin, U. Jansson, Synthesis and characterization of multicomponent (CrNbTaTiW)C films for increased hardness and corrosion resistance, Mater. Des. 149 (2018) 51–62. https://doi.org/10.1016/j.matdes.2018.03.068.

[51] I.M. Vinitskii, Relation between the properties of groups IV-V transition metals and carbon content, Poroshkovaya Metall. (1972) 76–82.

[52] M.D. Hossain, J.-P.M. Borman, Trent, Oses, Corey, C. Toher, S. Curtarolo, D. Brenner, Bipolar High Power Impulse Magnetron Sputtering of High Entropy Carbides, (n.d.).

[53] M. Samuelsson, K. Sarakinos, H. Högberg, E. Lewin, U. Jansson, B. Wälivaara, H. Ljungcrantz, U. Helmersson, Growth of Ti-C nanocomposite films by reactive high power impulse magnetron sputtering under industrial conditions, Surf. Coatings Technol. 206 (2012) 2396–2402. https://doi.org/10.1016/j.surfcoat.2011.10.039.

[54] X.X. Yu, G.B. Thompson, C.R. Weinberger, Influence of carbon vacancy formation on the elastic constants and hardening mechanisms in transition metal carbides, J. Eur. Ceram. Soc. 35 (2015) 95–103. https://doi.org/10.1016/j.jeurceramsoc.2014.08.021.

[55] R.H.J. Hannink, M.J. Murray, The effect of domain size on the hardness of ordered $VC_{0.84}$, Acta Metall. 20 (1972) 123–131. https://doi.org/10.1016/0001-6160(72)90120-4.

[56] G. Morgan, M.H. Lewis, Hardness anisotropy in niobium carbide, J. Mater. Sci. 9 (1974) 349–358.